# Theory on the mechanism of site-specific DNA-protein interactions in the presence of traps

*G. Niranjani and R. Murugan**

*Department of Biotechnology, Indian Institute of Technology Madras*
*Chennai, India*

* rmurugan@gmail.com

## ABSTRACT

The speed of site-specific binding of transcription factor (TFs) proteins with genomic DNA seems to be strongly retarded by the randomly occurring sequence traps. Traps are those DNA sequences sharing significant similarity with the original specific binding sites. It is an intriguing question how the naturally occurring TFs and their specific binding sites are designed to manage the retarding effects of such randomly occurring traps. We develop a simple random walk model on the site-specific binding of TFs with genomic DNA in the presence of sequence traps. Our dynamical model predicts that (a) the retarding effects of traps will be minimum when the traps are arranged around the specific binding site such that there is a negative correlation between the binding strength of TFs with traps and the distance of traps from the specific binding site and (b) the retarding effects of sequence traps can be appeased by the condensed conformational state of DNA. Our computational analysis results on the distribution of sequence traps around the putative binding sites of various TFs in mouse and human genome clearly agree well the theoretical predictions. We propose that the distribution of traps can be used as an additional metric to efficiently identify the specific binding sites of TFs on genomic DNA.

## KEY WORDS

## 1. Introduction

Binding of transcription factor proteins (TFs) at specific *cis*-regulatory modules (CRMs) on the genomic DNA in the presence of enormous amount non-specific sites is critical for the expression and regulation of various genes inside the living cell [1-5]. In earlier studies, the site-specific binding of TFs was modelled as one-step Smolochowski type three-dimensional diffusion (3Dd) controlled collision process [6]. Later *in vitro* experiments on the site-specific binding of *lac* repressor protein with its Operator site showed a bimolecular collision rate in the order of $10^{10}$-$10^{11}$ M$^{-1}$s$^{-1}$ which was 10-$10^2$ times higher than the Smolochowski type 3Dd controlled rate limit [6-8]. These experimental observations clearly ruled out the possibility of 3Dd only (3Ddo) model and suggested a two-step mechanism i.e. a 3D1Dd model. According to this, TFs first non-specifically bind with DNA via 3Dd and then search for their specific sites via one-dimensional diffusion (1Dd) [7, 8].

The nonspecific binding of TFs is mainly driven by the electrostatic attractive forces present in between the positively charged DNA binding domains (DBDs) of DNA binding proteins (DBPs) and negatively charged phosphate backbone of DNA. Various symbols and abbreviations used in this paper are summarized in **Table 1**. The site-specific binding of TFs via a combination of 3Dd and 1Dd seems (3D1Dd model) to be facilitated by sliding, hopping and inter-segmental transfer type dynamics [7-12]. The non-specifically bound TFs diffuse along DNA with a step size of unit base-pair (bps) in sliding, few bps in hopping and few hundred to thousand bps in intersegmental transfers. Intersegmental transfers occur when two distal segments of the same DNA polymer come into contact through ring-closure events over 3D space (**Fig. 1A**). The conformational state of DNA seems to play critical role in modulating various facilitating processes. Sliding and hopping will be the dominating modes of dynamics on a relaxed conformational state of DNA (rcsDNA). Intersegmental transfers predominantly occur on a condensed conformational state of DNA (ccsDNA).

In general, both the nonspecific and site-specific binding of TFs are influenced by several factors viz. conformational state of DNA [13-15], electrostatic attractive forces acting at the DNA-protein interface [16, 17] and the counteracting shielding effects of solvent ions [10], presence of semi-stationary [18] and dynamic roadblocks on DNA [19-21], conformational fluctuations in the DBDs of TFs [22, 23] (**Fig. 1B**), and randomly occurring kinetic traps along the DNA sequence [24-26]. Apart from these factors, the spatial organization of the genome structure also play important roles in accelerating the search process of TFs for their cognate sites on DNA [27, 28]. The electrostatic interactions along with the counteracting shielding effects of solvent ions creates a fluidic type environment for the 1Dd of DBDs of TFs at the DNA-protein interface.

Presence of roadblocks increases the dissociation of TFs and subsequently drive the mode of dynamics towards 3Dd mediated excursions [25, 26]. Switching between 1Dd and 3Dd will be enhanced by the conformational fluctuations at the DBDs of TFs across stationary and mobile states [22]. Here the stationary state is more sensitive to the sequence information than mobile one but moves slowly along DNA. Whereas mobile state is less sensitive to the sequence information but moves quickly along DNA. The conformational fluctuations in the DBDs of DBPs seem to be purely thermal driven. The free energy barrier that separates the stationary and mobile states seems to be close to the thermal energy which resembles the dynamics of downhill folding proteins at their mid-point denaturation temperatures [29]. Detailed calculations showed that the extent of thermodynamic coupling between the search dynamics and the rate of conformational fluctuations in the DBDs of TFs will be an optimum at the barrier of $\sim k_B T \ln 2$ [29] .

Presence of sequence mediated kinetic traps retards the rate of site-specific binding of TFs in several ways. First of all there is a nonzero probability of occurrence of trap sites similar to the **s**pecific **b**inding **s**ite (**SBS**) of TFs for sufficiently large genomes. Sequence traps slow down the 1Dd dynamics of TFs and also increase the overall dissociation time compared to that of other **n**onspecific **b**inding **s**ites (**NSBS**). In other words, kinetic traps increases the overall ruggedness or frustration of the binding energy landscape of the DNA sequence. It is still not clear how exactly the effects of kinetics traps are handled by various TFs under *in vivo* conditions. Recent experimental studies suggested that the presence of similar binding site adjacent to the specific binding site significantly influences the site-specific association rate [14, 24, 30]. Especially the extent of such influences seems to be directly proportional to the distance between the SBSs and the kinetic traps. In this context it is still not clear about (**a**) how the relative position of traps with respect to the position of initial nonspecific contact of TFs with DNA influences the site-specific association rate and (**b**) how the distribution of the distances between traps and SBSs in the real genomes (whether it is randomly distributed or correlated) affects the overall site-specific binding of TFs. Using a combination of theoretical and computational tools we will address these issues in detail.

## 2. Theory

Presence of sequence traps influences the binding of TFs with DNA in two different ways viz. (a) they compete with other **NSBS**s for the available pool of TFs and TFs bound at traps will be temporarily immobilized and (b) they retard the 1Dd dynamics of TFs on DNA. One can approximately ignore the competing effects of traps since the number of traps in a genome will be much lesser than the number of NSBSs. The overall search time or mean first passage time (MFPT) $\tau_{S,U}$ associated with TFs to find their SBSs on DNA can be written as follows [7, 8, 30, 31].

$$\tau_{S,U} \simeq \left[ \left[ P_{BTF} \right] \left( k_{fa} + k_{fX} \big/ \left( 1 + k_d \lambda \eta_U \right) \right) \right]^{-1} ; \; \lambda = N/U ; \; \eta_U = U^2/12D_o \qquad [1]$$

In this equation $P_{BTF}$ (M, mols/lit) is the concentration of TFs of interest in cytoplasm, $k_{fa}$ (M$^{-1}$ s$^{-1}$) is the bimolecular rate constant associated with the direct site-specific binding of TFs via 3Ddo mode, $k_{fX}$ is the overall non-specific binding rate and $k_d$ (s$^{-1}$) is the dissociation rate of nonspecifically bound TFs. Further $\lambda$ is the number of association-scan-dissociation (ASD) cycles required by TFs to scan the entire DNA and $\eta_U$ is the overall average time that is required by TFs to scan $U$ bps of DNA before dissociation where $U$ is a random variable that will take different values in each ASD cycle. The probability density function associated with the 1Dd lengths $U$ can be written as follows [31].

$$p_U \left( U \right) \simeq 2U e^{-\left( U/\Pi_A \right)^2} \big/ \Pi_A^2 ; \; \Pi_A = \sqrt{12D_0/k_d} \qquad [2]$$

Here $\Pi_A$ is the maximum possible 1Dd length of nonspecifically bound TFs on DNA that is measured in bps where 1bps ~ 3.4 x 10$^{-10}$ m and $D_o$ (bps$^2$s$^{-1}$) is the diffusion coefficient associated with the 1Dd of TFs on DNA. Clearly presence of sequence traps increases the 1Dd time $\eta_U$ which in turn increases the overall search time. In the following section we will try to understand the effects of sequence mediated kinetic traps on 1Dd dynamics of TFs within the framework of random walks with random hop size.

## 2.1. Random walks with random hop size and traps

Let us consider the DNA as a linear lattice confined within ($x_L$, $x_R$). Inside this lattice we consider DBP as an unbiased **1D r**andom **w**alker (**1Drw**) that is searching for the absorbing point at $x = x_A$ and ($x_L$, $x_R$) are the reflecting boundaries (**Fig. 2A**). Here the absorbing point is the CRMs (SBS) associated with the TFs which means that whenever TFs hit these points then transcription initiation starts approximately with a probability of one. The Langevin type stochastic differential equation that describes the dynamics of such 1Drw can be written as follows [32-35].

$$dx/dt = \sqrt{d_o}\,\Gamma_t;\ t = t_0;\ x = x_Z;\ \langle \Gamma_t \rangle = 0;\ \langle \Gamma_t \Gamma_{t'} \rangle = \delta\left(t - t'\right) \qquad [3]$$

Here $x$ is the position of 1Drw at time $t$ with the condition that it was at $x = x_Z$ at $t = t_0$ and $\Gamma_t$ is the Gaussian white noise whose mean and covariance properties are defined as in **Eq. 3**. From **Eq. 3** one can conclude about the mean and variance of the position of an unbiased 1Drw performing normal diffusion as follows.

$$x = x_Z + \sqrt{d_o}\int_0^t \Gamma_s ds;\ \langle x \rangle = x_Z;\ \langle x^2 \rangle - \langle x \rangle^2 = d_0 t \qquad [4]$$

Here $d_0$ is the phenomenological 1Dd coefficient. The probability density function associated with the dynamics of the 1Drw on a linear lattice obeys the Fokker-Planck equation (FPE) which can be written along with the boundary conditions as follows.

$$\partial_t P\left(x,t \mid x_Z,t_0\right) = \left(d_o/2\right)\partial_x^2 P\left(x,t \mid x_Z,t_0\right);\ \left[\partial_x P\left(x,t \mid x_Z,t_0\right)\right]_{x=x_L} = 0;\ P\left(x_A,t \mid x_Z,t_0\right) = 0 \qquad [5]$$

Here $P\left(x,t \mid x_Z,t_0\right)$ is the probability of observing 1Drw at position $x$ at time $t$ with the condition that it was at $x = x_Z$ at $t = t_0$. Apart from the boundary conditions given in **Eq. 5** we also set the initial condition as $P\left(x,t_0 \mid x_Z,t_0\right) = \delta\left(x - x_Z\right)$. When the 1Drw of interest moves with unit step size then the 1Dd coefficient can be defined as $d_o = \sum_{i=-l_d}^{l_d} p_i w_i i^2$ where $w_{\pm i}$ are the microscopic transition rates associated with the forward and reverse movements of 1Drw and $p_{\pm i}$ are the corresponding microscopic transition probabilities [10, 32]. In case of site-specific DNA protein interactions, the step length $i$ is measured in terms of base-pairs (bps). We have defined $l_d = 1$bps. Since the dynamics at the DNA-protein interface involves segmental motion of DBDs of TFs one can assume protein folding rate limit [36] for the transition rates as $w_{\pm i} \sim 10^6\,\text{s}^{-1}$. Noting that $p_{\pm i} \sim \frac{1}{2}$ for an unbiased 1Drw one finds that $d_o \sim 10^6\,\text{bps}^2\text{s}^{-1}$. Approximately this is the value of 1Dd coefficient associated with the sliding dynamics [19, 31, 37]. To simplify our analysis and other computations we use the following scaling transformations so that the dynamical variables in **Eq. 5** become dimensionless.

$$\langle w_{\pm i} \rangle = \phi;\ \tau = \phi t;\ X = x/l_d;\ D_o = d_o/\phi l_d^2 \qquad [6]$$

When $w_{\pm i} = \phi$ and $p_{\pm i} \sim \frac{1}{2}$ then $D_o = 1$. Upon rescaling the variables in **Eq. 5** as in **Eq. 6** we obtain the following Fokker-Planck equation (FPE) in dimensionless form.

$$\partial_\tau P\left(X,\tau \mid X_Z,\tau_0\right) = \left(D_o/2\right)\partial_X^2 P\left(X,\tau \mid X_Z,\tau_0\right) \qquad [7]$$

The corresponding initial and boundary conditions are as follows.

$$P\left(X, \tau_0 \mid X_Z, \tau_0\right) = \delta\left(X - X_Z\right); \; \left[\partial_X P\left(X, \tau \mid X_Z, \tau_0\right)\right]_{X=X_L} = 0; \; P\left(X_A, \tau \mid X_Z, \tau_0\right) = 0 \qquad [8]$$

The mean first passage time (MFPT) that is measured in terms of dimensionless number of steps associated with the escape of the 1Drw from the interval $(X_L, X_R)$ through the absorbing point $X_A = X_R$ starting from $X_Z$ will obey the following backward type FPE with appropriate boundary conditions.

$$D_o d_X^2 \Pi_S = -2; \; \left[d_X \Pi_S\right]_{X=X_L} = 0; \; \left[\Pi_S\right]_{X=X_R} = 0; \; \Pi_S = \left(X_R^2 - X_Z^2 - 2X_L\left(X_R - X_Z\right)\right)\Big/ D_o \qquad [9]$$

The results presented in **Eqs. 3-9** are standard and well known [9, 32, 38, 39]. Now we introduce traps at random locations inside the interval $(X_L, X_R)$. Traps are different from the absorbing boundaries in the sense that the free energy associated with the binding of TFs at traps will be much lower than SBSs but much higher than NSBSs. When traps act as sinks (absorbing boundaries) for the inflowing TFs then the overall probability $(p_{sp})$ associated with the nonspecifically bound TFs to specifically bind with their targets will be $p_{sp} < 1$ [24, 30, 40] which is mainly due to the partitioning of trajectories of inflowing TFs between SBSs and traps. In our realistic model, binding of TFs at traps cannot initiate transcription i.e. traps are neither sources nor sinks for the probability influx associated with TFs towards their SBSs.

However traps can significantly slowdown the site specific binding rate and hence the overall transcription rate will be reduced. In this background we assume that when 1Drw visits a trap, it will get stuck there for a fixed average amount of time (dwell time) and then escapes back into the original lattice interval. We denote the position of $r^{\text{th}}$ trap on linear lattice as $X_r$ where the subscript $r$ ranges from 1 to $m$ and the corresponding average dwell times of 1Drws at these traps are denoted as $\rho_r$. We denote the microscopic rate constant associated with the dissociation of TFs from a trap at position $X_r$ in a dimensionless form as $k_r/\phi$. The probability density function associated with the distribution of dwell times of 1Drw at the particular trap will be $p_r\left(\rho\right) = \left(k_r/\phi\right)\exp\left(-k_r\rho/\phi\right)$ [41, 42]. So that the mean dwell time associated with the 1Drw that was stuck at the respective trap that is present at the lattice position $X_r$ can be defined as $\rho_r = \int_0^\infty \rho\, p_r\left(\rho\right)d\rho$. This can be explicitly written as $\rho_r = \phi/k_r$. Here $k_r$ and $\phi$ are measured in s$^{-1}$ so that $\rho_r$ will be measured in dimensionless number of simulation steps since the average time required for each forward or reverse movement of 1Drw will be $1/\phi$.

There are $n$ number of traps in $(X_L, X_Z)$ and $m$-$n$ number of traps in $(X_Z, X_R)$ and there are totally $m$ number of traps inside the entire interval $(X_L, X_R)$. Here trap positions act a partial absorbing boundaries for the backward FPE defined in **Eq. 9**. The backward FPE given in **Eq. 9** cannot be solved analytically with so many number of boundary conditions. However one can derive an approximate formula for the overall MFPT in the presence of traps using the following arguments. First of all one should note that the MFPT in the presence of traps will be always higher than the MFPT in the absence of traps i.e. traps always retard the 1Dd of 1Drw. When $X_L = 0$ and $X_Z = 0$, then all the traps will be located inside $(X_Z, X_R)$ and from **Eq. 9** one finds that the random walker visits each and every site of the interval on an average $X_R$ number of times under such conditions. This means that the total dwell time of 1Drw at those $m$ number of traps that is added up to the original MFPT associated with the case of no traps in **Eq. 9** will be directly proportional to $\rho_T \propto \sum_{r=1}^{m} X_R \rho_r$. This result follows from the fact that when $X_Z = 0$, $X_L = 0$ and $k = 1$, then $D_o = 1$ and one obtains $\Pi_S = X_R^2$ from which one can conclude that 1Drw visits each site of the interval ($X_L = 0$, $X_R$) on an average $X_R$ number of times before reaching the absorbing point $X_R$.

Now let us assume that $X_Z$ is located well inside ($X_L$, $X_R$) such that $X_L < X_Z < X_R$ and $X_L = 0$. Depending on the relative position of the trap, the corresponding dwell time will vary. When there are $n$ number of traps with positions $X_r$ where $r = 1$ to $n$ located inside ($X_L$, $X_Z$) such that $X_L < X_r < X_Z < X_R$ then the total dwell time ($\rho_T$) will be directly proportional to the splitting probability associated with 1Drw to reach $X_L$ starting from $X_Z$ i.e. $\rho_T \propto \sum_{r=1}^{n} X_R \rho_r p_{LZ}$ where we have defined the splitting probability associated with the 1Drw to reach $X_L$ (assuming that $X_L$ is a probability sink) from $X_Z$ as $p_{LZ} = \left(1 - X_Z / X_R\right)$ [32]. This is because there is a nonzero probability for 1Drw to reach the absorbing point $X_A = X_R$ that is located right side of $X_Z$ without visiting the traps located in the left side of $X_Z$. On the other hand when there are $m$-$n$ number of traps located inside ($X_Z$, $X_R$) such that $X_L < X_Z < X_r < X_R$, then the total dwell time will be directly proportional to the splitting probabilities associated with 1Drw to reach $X_L$ starting from the respective trap positions $X_r$ as $\rho_T \propto \sum_{r=n+1}^{m} X_R \rho_r p_{Lr}$ which decides the number of times 1Drw revisits the corresponding traps located in ($X_Z$, $X_R$). Here the splitting probability associated with the 1Drw to reach $X_L$ starting from $X_r$ (assuming that $X_L$ is a probability sink) can be defined as $p_{Lr} = \left(1 - X_r / X_R\right)$. When $X_L = 0$, $X_A = X_R$ and noting the fact that $D_o = 1$ for the hop size of $k = 1$ then one can summarize these results using the following approximate formula.

$$\tilde{\Pi}_S \simeq \Pi_S + 2X_R \left\{ \begin{array}{l} \sum_{r=1}^{n} \rho_r \left(1 - X_Z / X_R\right) \mid X_L < X_r \leq X_Z \\ \sum_{r=n+1}^{m} \rho_r \left(1 - X_r / X_R\right) \mid X_Z \leq X_r < X_R \end{array} \right\}; \ \Pi_S = \left(X_R^2 - X_Z^2\right) \tag{10}$$

Here the factor 2 appears in the expression for dwell time since the flux of those trajectories flowing towards $X_L$ will be again reflected back towards the absorbing boundary $X_R$. Now we drop the idea of unit step size movement of 1Drw and assume that the random walker can hop for $k$ steps at a time. Suppose the current position of 1Drw is $X$. Then in the next step it can move anywhere in [$X$-$k$, $X$+$k$] with equal probabilities i.e. $1/2k$ and one finds that $d_o = \sum_{i=-kl_d}^{kl_d} p_i w_i i^2$. Though we have $d_o \propto i^2$, 1Dd coefficient cannot be higher than 3Dd coefficient ($d_t$) since there is a strong negative correlation between $p_{\pm i}$ and $i$. In general $d_o \leq d_t$ [31]. In the context of site-specific DNA-protein interactions, the hop size is directly proportional to the degree of condensation or supercoiling of the DNA polymer. This means that ccsDNA favors higher hop sizes than rcsDNA. Following the detailed theory of random walks with random step size [38, 39] one can derive the following expression for the MFPT for 1Drw with random hop size of $k$.

$$\tilde{\Pi}_S \simeq \Pi_S + \frac{2X_R}{D_o} \left\{ \begin{array}{l} \sum_{r=1}^{n} \rho_r p_{LZ} \mid X_L < X_r \leq X_Z \\ \sum_{r=n+1}^{m} \rho_r p_{Lr} \mid X_Z \leq X_r < X_R \end{array} \right\} + 2\left(X_R - X_L + \sum_{t=1}^{m} \rho_r\right)\left(1 - 1/k\right) \tag{11}$$

For $X_L \leq X_Z < X_A = X_R$, when transition rates are $w_{\pm i} = \phi$ and the transition probabilities are $p_{\pm i} \simeq 1/2k$ then one finds that $D_o \simeq (k+1)(2k+1)/6$. Clearly for the hop size of $k = 1$, we find that $D_o = 1$ and **Eq. 11** reduces to **Eq. 10**. When $k > 1$; $X_L = 0$; $X_A = X_R$ and $X_L < X_t \leq X_Z$ or $X_Z \leq X_t < X_R$ then one can derive the following explicit expression for the overall MFPT associated with the escape of 1Drw through $X_R$ starting from $X_Z$ in the presence of $m$ traps.

$$\tilde{\Pi}_S \simeq \left(X_R^2 - X_Z^2\right)\Big/D_o + \frac{2X_R}{D_o}\left\{ \begin{array}{l} \sum_{r=1}^{n} \rho_r p_{LZ} \mid X_L < X_r \leq X_Z \\ \sum_{r=n+1}^{m} \rho_r p_{Lr} \mid X_Z \leq X_r < X_R \end{array} \right\} + 2\left(X_R + \sum_{t=1}^{m}\rho_r\right)(1-1/k) \qquad [12]$$

Here one can define the initial position as well as trap position averaged MFPT associated with 1Drw to reach $X_R$ starting from $X_Z$ as follows i.e. $\overline{\Pi}_S = \left(\int_0^{X_R} \tilde{\Pi}_S dX_Z + \int_0^{X_R} \tilde{\Pi}_S dX_r\right)\Big/X_R$.

$$\overline{\Pi}_S = 6\left(2X_R^2/3 + X_R\sum_{r=1}^{m}\rho_r\right)\Big/(k+1)(2k+1) + 2\left(X_R + \sum_{r=1}^{m}\rho_r\right)(1-1/k) \qquad [13]$$

For $k = 1$ and sufficiently large values of hop size $k$, **Eq. 13** suggests the following limiting conditions.

$$\lim_{k \to 1} \overline{\Pi}_S = 2X_R^2/3 + X_R\sum_{r=1}^{m}\rho_r; \quad \lim_{k \to \infty} \overline{\Pi}_S = 2\left(X_R + \sum_{r=1}^{m}\rho_r\right) \qquad [14]$$

**Eq. 14** suggests that at sufficiently large values of hop size 1Drw visits each sites of the linear lattice for only once. One needs to multiply $\overline{\Pi}_S$ by $\phi$ to transform it back into time units in seconds where $X_R$ will be still a dimensionless count as in **Eq. 6**. **Eqs. 10-14** are the central results of this paper which clearly suggest that in the presence of sequence traps the 1Dd time $\eta_U$ will transform as $\eta_U \mapsto \eta_U + H_\rho$ where $H_\rho$ is the sum of contributions from the dwell times of TFs at sequence traps.

## 3. Results and Discussion
### 3.1. Effects of sequence traps on 1Dd time of TFs
From **Eq. 10** one can conclude that the overall delay in the search time $\eta_U$ due to the presence of traps will be (**a**) directly proportional to the distance of the initial position of TFs from the absorbing boundary located at SBSs ($X_R$-$X_Z$) when the traps are not located in between ($X_Z$, $X_R$), (**b**) directly proportional to the distances of the traps from the SBSs ($X_R$-$X_r$) when the traps are located in between $X_Z$ and $X_R$. Further (**c**) relatively less delay in the search time due to the traps will be possible only when there is a negative correlation between the dwell times $\rho_r$ and the corresponding distances of traps ($X_R$-$X_r$) from the SBSs which ensure that the second sum in **Eq. 10** will be at minimum. Here observation (**c**) is an important one which tell us about the principles associated with the design of CRMs. That is to say the efficiency of CRMs in the gene regulation can be fine-tuned by suitably arranging a set of traps around them. In other words a set of identical CRMs can be well differentiated by appropriately arranging the kinetic traps around them. In the following sections we will first check the validity of the approximate expressions given by **Eq. 10** and **12** by detailed random walk simulations. Then we will analyze the distribution of naturally occurring SBSs of various TFs and the associated sequence traps to check the validity of our propositions **a – c**.

### 3.2. Stochastic simulations on random walks with random hop size and traps

To check the validity of **Eqs. 11-12** we performed stochastic random walk simulations. We considered an unbiased 1Drw on a linear lattice of length with $X_R = 25$ and $X_L = 0$. We first simulated with hop size of $k = 1$. Here $X_L$ is a reflecting boundary and $X_A = X_R$ is the absorbing boundary. This means that when 1Drw whose current position is $X = 0$ tries to visit $X = -1$ then it will be put pack to $X = 1$. With these settings when $X_Z = 0$ then from **Eq. 10** one obtains $\Pi_S = 625$ since $D_o = 1$. When $X_Z > 0$, then the MFPT associated with the escape of 1Drw into $X_R$ starting from $X_Z$ will be given by $\Pi_S = 625 - X_Z^2$ which is a well-known result. We measure MFPT in terms of dimensionless number of simulation steps taken by 1Drw to reach $X_A$ starting from $X_Z$.

Now we introduce a single trap at $X_r$ and fix the initial position arbitrarily at $X_Z = 13$. The trap at $X_r$ is characterized by a dwell time of $\rho_r = 25$ simulation steps. This means that whenever 1Drw visits $X = X_r$ then it will get stuck there for 25 simulation steps. With this settings we iterated $X_r$ from 1 to 24 and the results are shown in **Fig. 2B**. Clearly when $X_r < X_Z = 13$ then using **Eq. 11** one can compute the MFPT as $\Pi_S = 456 + 2\rho_r \left( X_R - X_Z \right) = 1056$. When $X_r > X_Z$ then one finds that $\Pi_S = 456 + 50\left( 25 - X_r \right)$. With these settings we increase the hop size from 1 to 2, 3 and 4. Under such conditions one finds that $D_o = \left( k + 1 \right)\left( 2k + 1 \right)/6$. When $X_r < 13$ then using **Eq. 12** one can compute the MFPT for $k = 2$ as $\Pi_S \simeq 472$. For $k = 3$ one finds that $\Pi_S \simeq 293$ and for $k = 4$ we find that $\Pi_S \simeq 216$. The simulation results depicted in **Fig. 2C** are consistent with these theoretical predictions by **Eq. 10** for $m = 1$.

To check the validity of **Eq. 10** for many number of traps i.e. $m > 1$ we introduced traps at arbitrary locations $X_r = [3, 5, 8, 11, 15, 18, 21, 24]$ i.e. $m = 8$. The corresponding dwell times for these traps were set at $\rho_r = [5, 15, 25, 35, 45, 55, 65, 75]$ and iterated $X_Z$ from 1 to 24. This means that $n$ will vary depending on $X_Z$. For $X_Z = 1$ one finds that $n = 0$, for $X_Z = 12$ one finds that $n = 4$ and so on. The simulation results for $X_L = 0$, $X_A = X_R = 25$ are shown in **Fig. 2C** which agree well with the theoretical predictions by **Eq. 10**. The effects of various types of correlation between the variables $\left( X_R - X_r, \rho_r \right)$ are shown in **Figs. 2C**, **D** and **E**. Here minimum value of MFPT can occur only when there is a negative correlation between these variables as shown in **Fig. 2C**. The MFPT will be a maximum when there is a positive correlation between these variables (**Fig. 2D**). The MFPT corresponding to the case of random arrangement of dwell times will be somewhere in between these two extreme cases (**Fig. 2E**). **Figs. 2F-G** shows the effects of variation in $X_R$ and $X_Z$ over the MFPT associated with 1Drw to escape into $X_R$ starting from $X_Z$ in the presence of $m$ number of traps and various hop sizes $k > 1$. These stochastic random walk simulation results are consistent with the theoretical predictions by **Eqs. 10-11** which suggest that the overall MFPT will be insensitive to the variations of $X_Z$ at higher hop sizes.

### 3.3. Naturally occurring TF binding sites and their traps

Let us consider a genomic DNA of size $N$ bps containing a SBS (CRM) on it for a TF protein of interest. We denote the binding stretch of TFs as $q$ bps and assume that the probability associated with the occurrence of each base A, T, G and C in the genome is ¼. With this setting one can calculate the minimum required value of $q = q_C$ to ensure that there is only one copy of the SBS in the entire genome by chance as follows.

$$s = \left(N - q\right)\left(1/4\right)^q \; ; \; s - 1 = 0; \; q_C = \left(2N - \mathrm{LambertW}\left(2^{2N+1}\ln 2\right)\right)\big/2\ln 2 \qquad [15]$$

In this equation $s$ is the number of similar binding sites with size of $q$ bps which can occur by chance and $y = \mathrm{LambertW}(x)$ is the solution of $y \exp(y) = x$. When $q < q_C$ for a transcription factor binding site then there is a definite chance of random occurrence of similar binding sites on the genome. One should also note the asymptotic result for a sufficiently large value of genome size $N$ as $q_C \sim \ln N/2\ln 2$. Size of haploid human and mouse genomes seem to be ~3.3 x $10^9$ and ~2.8 x $10^9$ bps respectively [2, 43, 44]. The corresponding critical length of binding stretch $q_C$ calculated from **Eq. 16** for human and mouse genomes will be ~15.8 and ~15.7 bps respectively (**Figs. 3A-B**). Statistical analysis of the binding stretches of TFs in human and mouse obtained from JASPAR database [45] suggested the mean values of $q$ as ~13 ± 0.43 and ~12.77 ± 0.43 bps respectively at a confidence level of 0.95. The median of the length of binding stretch of TFs in human seems to slightly higher (~13 bps) than the case of TFs in mouse (~12 bps). These results clearly suggest that $q < q_C$ for the genomes of human and mouse. This means that there is a nonzero probability of random occurrence of binding sites similar to that of SBSs corresponding to TFs on the same genomic sequence.

Using **Eq. 11** we have already shown that the delay in the overall search time due to the presence of traps will be minimum when there is a negative correlation between the dwell times $\rho_r$ and the corresponding distances of traps ($X_R$-$X_r$) from SBSs. This in turn predicts that an efficient configuration of the SBSs with traps should be such that those traps with high affinity for TFs should be closer to the original SBS than those traps with low affinity towards TFs. To check the validity of this prediction in natural system we computed the distribution of distances of traps with various binding affinities from the putative SBSs for various TFs in the upstream sequences of various genes of mouse and human.

The upstream 5000 bps sequences of various genes of human and mouse genomes were obtained from UCSC genome database (February 2009 assembly, hg19 version for human genome and December 2011 assembly, mm10 version of mouse genome) and position weight matrices (PWMs) [46, 47] of various TFs of mouse and human were obtained from publicly available JASPAR database [45, 48]. There were 28365 sequences from mouse genome and 28824 sequences from human genome. Using the PWMs of various TFs we generated the score table for various upstream sequences based on the following equation [46].

$$S_{v,i} = -\sum_{w=i+1}^{q}\left(\sum_{b=A}^{T} f_{b,w}\log_e\left(f_{b,w}/f_b\right)\right); \; b = \{A, C, G, T\} \qquad [16]$$

In this equation $S_{v,i}$ is the score value of PWM at $i$th position on $v$th sequence, $q$ is the length of binding stretch of the corresponding TF, $f_b$ is the background probability of observing base $b$ in the corresponding genome, and $f_{b,w}$ is the probability of observing base $b$ at position $w$ of the TF binding site. Here $f_b$ can be calculated from the random sequences corresponding to the given genome. There is a strong positive correlation between the score value and the binding free energy of TFs [46]. In parallel we also generated score table for random sequences using the same PWM from which we obtained the score distribution and the cutoff score value for the given matrix corresponding to a given p-value. In our calculations we have set the p-value < $10^{-6}$ for putative SBSs. Sample score table associated with the upstream sequence of mouse Fibin gene is depicted in **Fig. 3C** and the corresponding score distribution is shown in **Fig. 3D**. Here we have used the PWM corresponding to the mouse TF protein POU2F1a. We defined those lattice positions with score values higher than the cutoff score value with respect to p-value $< 10^{-6}$ as putative SBSs of the corresponding TFs. We further defined those sites whose score values corresponding to the p-values ranging from $10^{-4}$ to $10^{-6}$ as kinetic traps associated with the SBSs.

Based on these definitions we computed the probable distances of traps from the putative SBSs of various TFs. We combined these distances for various TFs and generated the overall histograms for the entire genome with a bin size of ~200 bps. We used the random sequences associated with each genome that is available at UCSC database to compute the probability of occurrence of putative binding sites by chance. We considered random sequences of size 5 x $10^6$ bps and fragmented it into $10^3$ number of sequences with length of 5000 bps. Then we scanned each random sequence with same PWM and obtained the number of putative SBSs (false positives). The probability of observing a SBS by chance will be calculated as $p_{NF}$ = number of false positives / 1000.

**Figs. 4A** and **B** shows the distribution of distances of traps from the putative SBSs where trap is defined by p-value $< 10^{-3}$ for human and mouse respectively. **Figs. 4C-D** shows the distribution of the distances of traps from SBSs where the trap is defined by p-value $< 10^{-4}$ and **Figs. 4E-F** correspond to those traps which are defined by p-value $< 10^{-5}$. We have defined those sites with p-values $< 10^{-6}$ as the putative SBSs. These results clearly suggest that those traps with strong affinity for TFs are located close to the transcription factor binding sites (TBS) in line with the prediction by **Eq. 11**.

Here one should note from **Fig. 3C** that apart from the putative SBSs and traps there are background fluctuations in the score values. Since score values are positively correlated with the binding energy, these background fluctuations form an integral part of the sequence dependent binding energy landscape for the TFs and upstream sequence of interest. A strong frustration in the free energy profile of a sequence significantly influences the 1Dd dynamics of TFs since the microscopic transition rates $w_{\pm i}$ in the definition of 1Dd are strongly dependent on the ruggedness of the binding energy landscape. Earlier studies suggested that $D_o \simeq D_o^0 \exp\left(-\varepsilon^2\right)$ [49-51] where $D_o^0$ is the 1Dd coefficient corresponding to a smooth free energy landscape and $\varepsilon$ is the degree of frustration or ruggedness (measured in $k_BT$) in the free energy profile of the DNA sequence. In this context one should note that the presence of traps further increases the degree of frustration in the binding free energy profiles of DNA sequence. From **Eqs. 10-11** we found that the retarding effects of such rough sequence potentials can be well handled by increasing the hop size associated with the 1Dd dynamics of TFs that in turn requires a ccsDNA [31].

### 3.4. Effects of traps which are identical to SBSs
In this section we consider a situation where there are two identical SBSs viz. CRM1 and CRM2 corresponding to the same TF (1Drw) present on the DNA polymer with a distance of $u$ bps from each other as shown in **Fig. 5** [30, 40]. When the mode of site-specific binding of corresponding TFs is via pure 3Dd route then the overall specific binding rate will be independent of the distance between those SBSs. Experimental results clearly corroborated the strong dependency of the overall site-specific binding rate on $u$ [30]. Since both are SBSs for the same TFs, the dwell time of TFs at these binding sites will be significantly high. Therefore each site can also generate a potential roadblock for the 1Dd movement of TFs towards the other site rather than merely acting as kinetic traps. This will clearly decreases the degrees of freedom associated with the approach of TFs towards the SBSs from left and right sides of CRMs via 1Dd from 2 to 1. As a consequence the influx of TFs especially from the right side of CRM1 as well as left side of CRM2 will be blocked by the TFs which are already bound on the respective CRMs. This will in turn decreases the effective number of SBSs from 2 to 1 especially when $u = 0$. This is because the probability of TFs visiting CRM1 from left side ($p_L$) will be ½ and the probability of TFs entering into CRM2 from the right side ($p_R$) will be ½ so that the total probability flux will be 1 since the probability flux through the interconnecting region is zero i.e. $p(0) = 0$.

Now we consider the influx of TFs through the interconnecting region between CRM1 and CRM2. Let us denote the average 1Dd length of TFs on DNA as $u_0$. One can compute the splitting probabilities associated with the TFs landing at the interconnecting region to reach any of the CRMs only when $u$ is comparable or less than that of $u_0$. When $u$ is much higher than $u_0$ then there is always a nonzero probability of TFs to dissociate from interconnecting region without meeting any of the CRMs. Alternatively the extent of roadblock will decrease as $u$ increases towards infinity which clearly uncouples these SBSs. In other words the incremental change in the probability of a TF which landed at the interconnecting region $p(u)$ to reach any one of the SBS with respect to $u$ will be (**a**) inversely proportional to $u$ and (**b**) directly proportional to the product of probabilities $p(u)\big(1 - p(u)\big)$. This is because when $u$ increases, then the extent of dynamic coupling of these CRMs via 1Dd of TFs decreases. When $u = u_0$ then the average (over the landing or initial position) splitting probabilities associated with the TFs landing at the interconnecting region to reach any one of the SBSs will be $p(u_0) = ½$. From these observations one can derive the following result.

$$dp(u)/du = p(u)\big(1 - p(u)\big)/u; \; p(u_0) = 1/2; \; \therefore \; p(u) = u/(u_0 + u) \tag{17}$$

From **Eq. 17** one can conclude that the total number of effective binding sites as seen by TFs from bulk will be $b(u) = p_L + p(u) + p_R$. Clearly $b(0) = 1$, $b(u_0) = 3/2$ and $b(\infty) = 2$. These results are consistent with the experimental observations [30, 40]. These results suggest that the presence of similar binding sites adjacent to SBSs significantly retards the 1Dd dynamics of TFs mainly via introducing stationary roadblocks rather than transiently trapping TFs.

## 4. Conclusions

In this paper we have investigated the effects of sequence traps on the site-specific binding of TFs at their respective *cis*-regulatory modules. We have shown that the speed of site-specific binding of TFs with DNA seems to be strongly retarded by the randomly occurring sequence traps. We developed a simple random walk model on the site-specific binding of TFs with genomic DNA in the presence of sequence traps. Our dynamical model predicted that (a) the retarding effects of traps will be minimum when the traps are arranged around the specific binding site such that there is a negative correlation between the binding strength of TFs with traps and the distance of traps from the specific binding site and (b) the retarding effects of sequence traps can be soothed by the condensed conformational state of genomic DNA. Our computational analysis results on the distribution of sequence traps around the putative binding sites of various TFs in mouse and human genome clearly agree well the theoretical predictions. We proposed that the distribution of traps can be used as an additional metric to efficiently identify the specific binding sites of TFs. It seems that the presence of similar binding sites adjacent to specific binding sites significantly retards the 1Dd dynamics of TFs mainly via introducing stationary roadblocks rather than transiently trapping TFs.

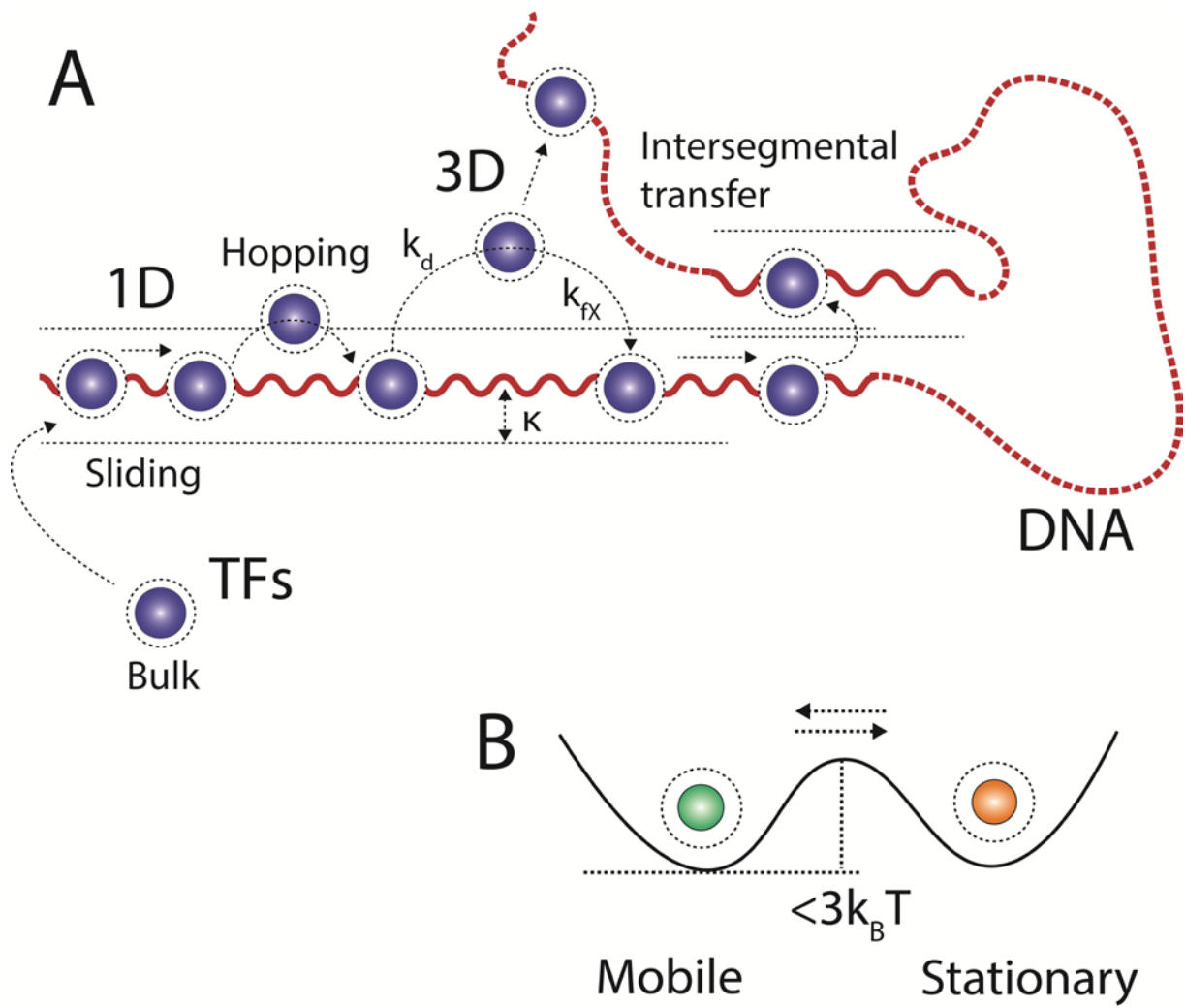

**FIGURE 1. A.** Various 1D facilitating processes involved in the site-specific binding of TFs with DNA. Here $k_{fX}$ (M⁻¹s⁻¹) is the bimolecular rate constant associated with the non-specific binding and $k_d$ (s⁻¹) is the rate constant associated with the dissociation of non-specific complex. Sliding is characterized by unit base-pair step size. Hopping involves few bps and intersegmental occurs when two distal segments of the same DNA polymer come close over 3D space via ring closure events. In all these sliding, hooping and intersegmental transfers TFs moves well within the Onsager radius ($\kappa$, measured in bps). Onsager radius is defined as the distance between the positively charged DNA binding domains of DBPs and negatively charged phosphate backbone of DNA at which the magnitude of the overall electrostatic attractive forces along with the counteracting shielding effects of solvent ions is equal to the background thermal energy. When TFs escape out the Onsager radius then we consider that as dissociation and subsequent 3D excursion. **B.** While searching for the cognate site, the DNA binding domains of TFs fluctuates between stationary and mobile conformations. Upon finding the specific site these fluctuations damps out to form a tight stationary state site-specific complex. This conformational flipping dynamics resembles as that of the downhill folding proteins at their mid-point denaturation temperatures.

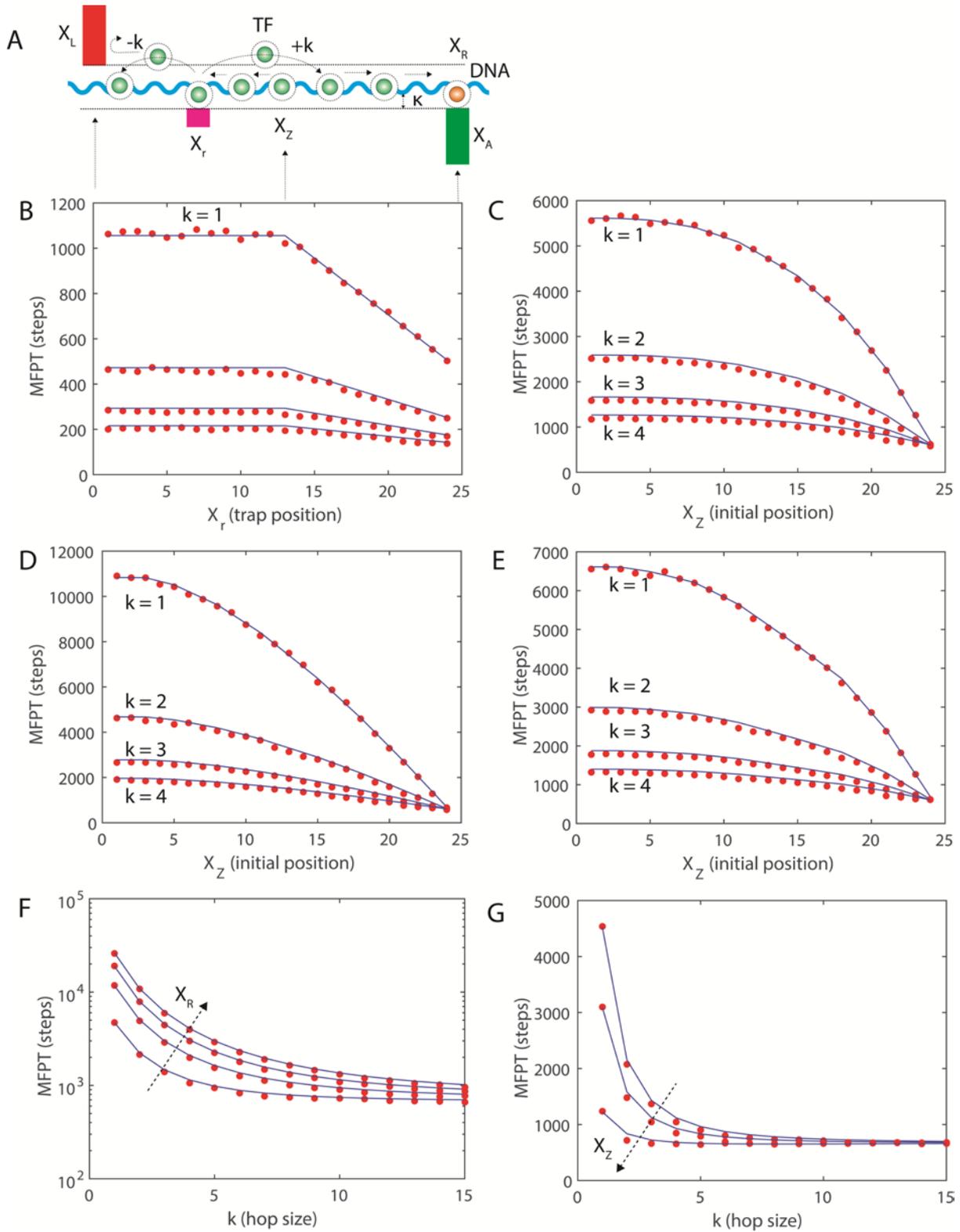

**FIGURE 2**: **A**. Random walk model with random hop size and traps. We consider a linear lattice confined in $(X_L, X_R)$ which are reflecting boundaries i.e. the random walker cannot escape over these boundaries. Inside this interval there is a 1Drw (TFs) searching for the absorbing point $X_A$ starting from $X_Z$. When 1Drw hits $X_A$ (SBS) then it will be removed from the system. With this setting, we introduce a trap at $X_r$. When 1Drw hits this trap then it will get stuck there for an average dwell time of $\rho_r$ and then escapes back into the original lattice.

In the context of site-specific DNA-protein interactions 1Drw will resemble TFs and absorbing point is the corresponding TF binding site (TBS). Here $\kappa$ is the Onsager radius. **B**. Here the settings are $X_L = 0$, $X_R = X_A = 25$, $\rho_r = 15$, $X_Z = 13$ and $X_r$ was iterated from 1 to 24. MFPT was computed over $10^5$ trajectories (red filled circles) and blue solid line is the prediction by **Eqs. 10-11** in all the simulations. **C**. Settings are $X_L = 0$, $X_R = X_A = 25$ and traps were arbitrarily kept at $X_r = [3, 5, 8, 11, 15, 18, 21, 24]$ and the corresponding $\rho_r = [5, 15, 25, 35, 45, 55, 65, 75]$ and $X_Z$ was iterated from 1 to 24. **D**. Settings are $X_L = 0$, $X_R = X_A = 25$ and traps were arbitrarily kept at $X_r = [3, 5, 8, 11, 15, 18, 21, 24]$ and the corresponding $\rho_r = [75, 65, 55, 45, 35, 25, 15, 5]$ and $X_Z$ was iterated from 1 to 24. **E**. Settings are $X_L = 0$, $X_R = X_A = 25$ and traps were arbitrarily kept at $X_r = [3, 5, 8, 11, 15, 18, 21, 24]$ and the corresponding $\rho_r = [25, 15, 45, 35, 5, 75, 65, 55]$ and $X_Z$ was iterated from 1 to 24. **F**. Settings are $X_L = 0$, $X_Z = 13$ and traps were arbitrarily kept at $X_r = [3, 5, 8, 11, 15, 18, 21, 24]$ and the corresponding $\rho_r = [5, 15, 25, 35, 45, 55, 65, 75]$ and the hop size $k$ was iterated from 1 to 15 and $X_R$ was iterated in $[25, 35, 45, 55]$ along the dotted arrow. **G**. Settings are $X_L = 0$, $X_R = X_A = 25$, $X_Z$ was iterated in $[14, 19, 23]$ along the dotted arrow and traps were at $X_r = [3, 5, 8, 11, 15, 18, 21, 24]$ with corresponding $\rho_r = [5, 15, 25, 35, 45, 55, 65, 75]$ and the hop size $k$ was iterated from 1 to 15.

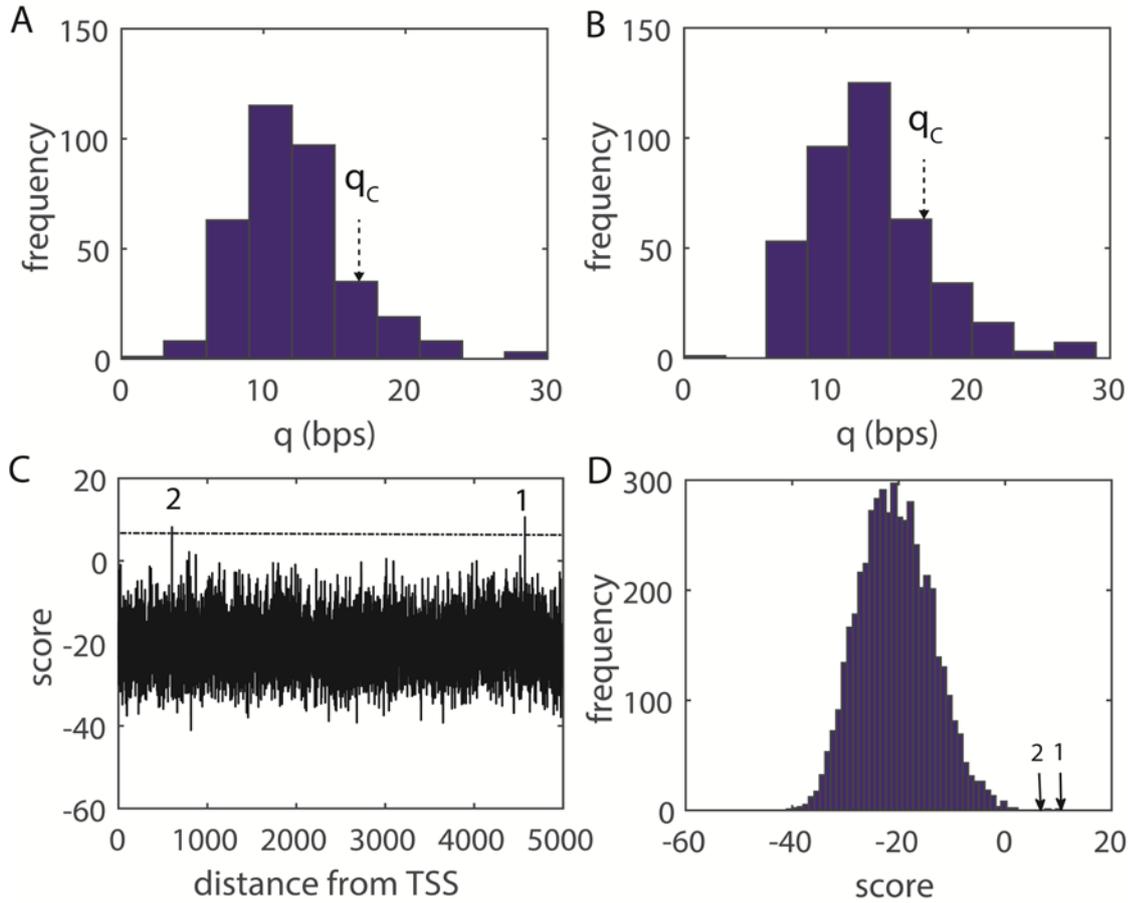

**FIGURE 3**: Distribution of binding stretch length ($q$) of TFs in mouse (**A**) and human (**B**). Computational analysis of PWMs from JASPAR database reveals that the mean values of $q$ as ~13±0.43 and ~12.77±0.43 bps respectively at a confidence level of 0.95. The median of the length of binding stretch of TFs in human seems to slightly higher (~13 bps) than the case of TFs in mouse (~12 bps). These results clearly suggest that $q < q_C$ for the genomes of human and mouse. **C**. Score table for the PWM corresponding to the mouse TF protein POU2F1a (NM_011137) that is scanned over *Mus musculus* fin bud initiation factor homolog gene (NM_026271). Here binding stretch length is $q = 18$ bps. The cutoff score for the definition of putative specific binding site with p-value $< 10^{-6}$ seems to be 10.16. Cutoff score for the definition of trap with p-value $< 10^{-5}$ seems to be 7.71 for $10^{-5}$. Here the **mark 1** is with specific binding site score of 10.58 which is 4570 bps away from transcription start site (TSS) and **mark 2** is the trap with score 8.19 which is 602 bps away from TSS and 3968 bps away from specific binding site. **D**. Distribution of score values obtained in **C** as defined by **Eq. 16**. The probability of observing a specific binding site by chance was $p_{NF} < 0.005$.

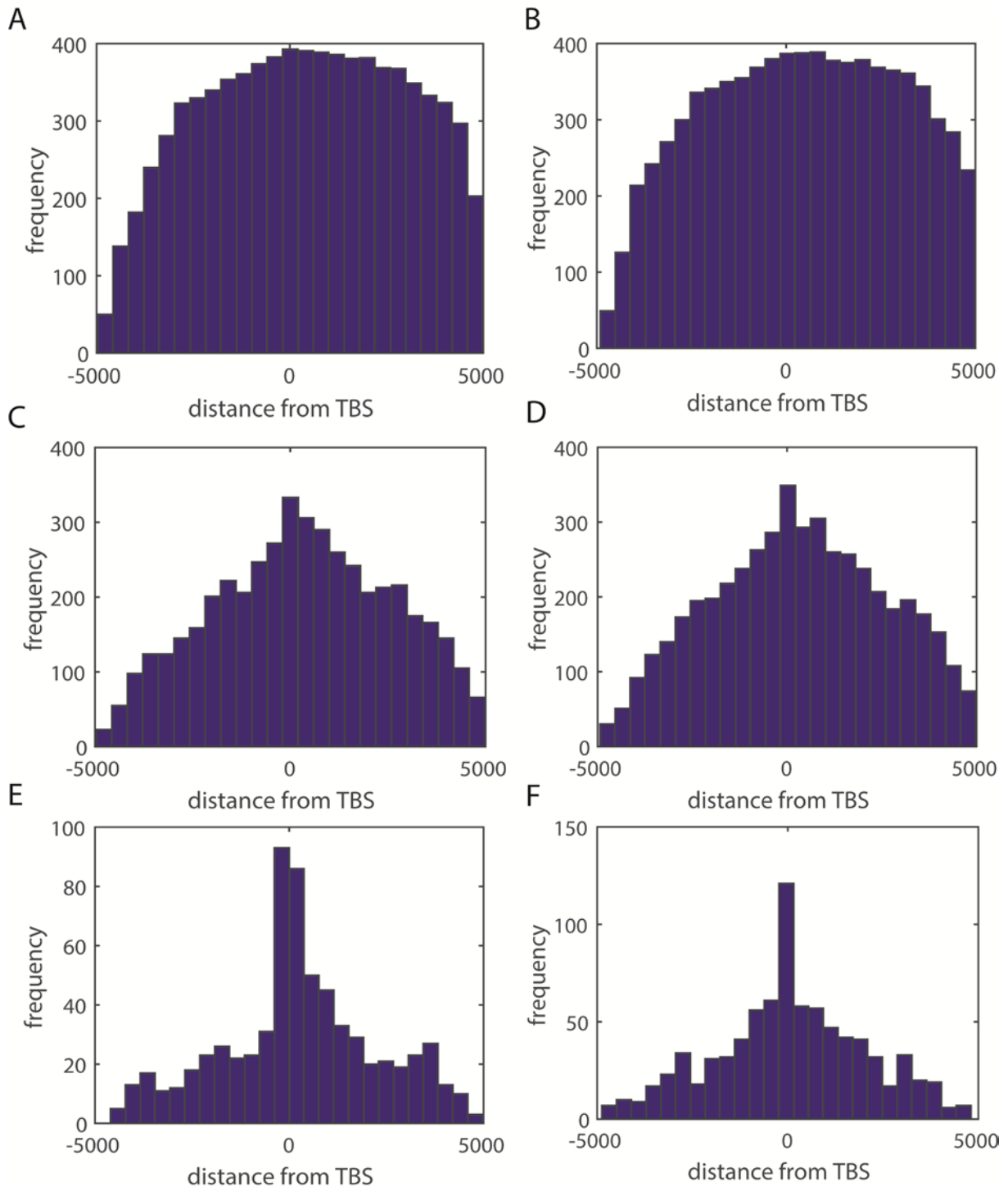

**FIGURE 4**. Distribution of distances between traps and putative specific binding sites in human (**B**, **D**, **F**) and mouse (**A**, **C**, **E**). These distances were obtained by scanning the upstream 5000 bps sequences of various genes of human and mouse with PWMs available for various human and mouse TFs in JASPAR database. For each combination of PWM and upstream sequence a score table was constructed and the cutoff scores for specific and trap sites were computed. In all these cases the specific binding site was defined by the p-value < $10^{-6}$. In **A** and **B**, the traps were defined by p-value < $10^{-3}$. In **C** and **D**, the traps were defined by p-value < $10^{-4}$. In **E** and **F** the traps were defined by p-value < $10^{-5}$. These results clearly suggest that traps with strong affinity towards TFs are preferably located near the specific binding site which is in line with the prediction of **Eqs. 10-11**.

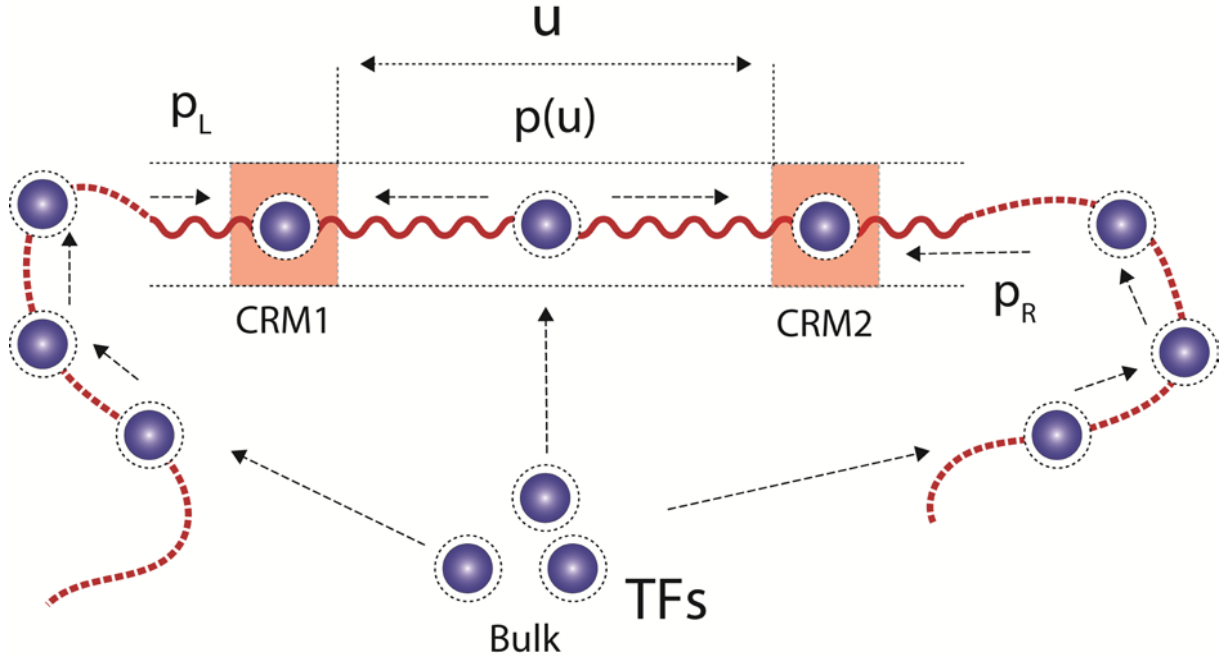

**FIGURE 5**: Effects of identical binding sites. In this model two identical specific binding sites CRM1 and CRM2 for the same TF was constructed with varying $u$ bps distance between them [30, 40]. Depending on $u$, the influence of CRM1 on CRM2 will vary. Clearly as $u$ tends towards infinity, both these binding sites behaves independently and the number of effective binding sites will be 2. When $u = 0$, then the number of effective binding sites will be 1 since $p_L = p_R = \frac{1}{2}$ and $p(0) = 0$. Here $p_L$ is the probability of TFs entering from left side of CRM1 and $p_R$ is the probability of TFs entering from right side of CRM2 and $p(u) = u/(u_0 + u)$ is the probability of TFs entering from the interconnecting region which can bind with any one of the CRMs as defined in **Eq. 17**. When $u = u_0$ where $u_0$ in the average 1Dd length of TFs, then the probability of TFs to reach any one of the CRMs via interconnecting region will be $p(u_0) = \frac{1}{2}$ so that the number of effective binding sites will be 3/2.

**Table 1. Various symbols and parameters used in the main text**

| Symbol | Definition | Remarks |
|---|---|---|
| SBS | Specific binding site | |
| NSBS | Nonspecific binding site | |
| DBD | DNA binding domain | |
| DBP | DNA binding protein | |
| TBS | Transcription factor binding site | |
| TSS | Transcription start site | |
| TF | Transcription factor | |
| nDd | n-dimensional diffusion | |
| CRM | cis-regulatory module | |
| rcsDNA | Relaxed conformational state of DNA | |
| ccsDNA | Condensed conformational state of DNA | |
| 3Ddo model | 3Dd only model on site-specific DNA-protein interactions in which the protein molecule will be assumed to bind at specific site on DNA via one-step 3Dd mediated routes. | |
| 3D1D model | In this two-step model, the protein binds at the specific site on DNA via a combination of 3Dd and 1Dd mediated routes. | |
| 1Drw | 1D random walker whose position on the linear lattice is x that is confined in $(x_L, x_R)$. | |
| $x_L$ | Left side boundary of linear lattice for 1Drw | bps |
| $x_R$ | Right side boundary of linear lattice for 1Drw | bps |
| $x_A$ | Absorbing boundary of linear lattice for 1Drw | bps |
| $x_Z$ | Initial position of 1Drw on linear lattice | bps |
| $x_r$ | Position of rth trap on linear lattice | |
| $l_d$ | base-pair, bps (1 bps = 3.4 x $10^{-10}$m) | |
| $X_L$ | $= x_L/l_d$ | dimensionless |
| $X_R$ | $= x_R/l_d$ | dimensionless |
| $X_A$ | $= x_A/l_d$ | dimensionless |
| $X_Z$ | $= x_Z/l_d$ | dimensionless |
| $X_r$ | $= x_r/l_d$ | dimensionless |
| $\phi$ | Average value of microscopic transition rates $<w_{\pm i}>$ over various hop sizes. | $s^{-1}$ |
| $k_r$ | Dissociation rate constant associated with the 1Drw that got stuck at rth trap. | $s^{-1}$ |
| $\rho_r$ | $(= \phi/k_r)$ dimensionless dwell time of 1Drw at rth trap | simulation steps |
| $k_{fX}$ | Rate constant associated with nonspecific contact formation between DNA and TFs | $M^{-1}s^{-1}$ |
| $k_d$ | Dissociation rate constant associated with the nonspecific DNA-protein complex. | $s^{-1}$ |
| $\kappa$ | Onsager radius which is defined as the distance between the DBDs of TFs and the phosphate backbone of DNA at which the overall electrostatic attractive forces will be comparable with that of the background thermal energy (~1$k_B$T). | bps |
| $k$ | Hop size associated with 1Drw or the DBPs which are | dimensionless |
| | performing 1Dd along DNA lattice. | |
| $d_o$ | 1Dd coefficient associated with the DBPs. In case of 1Drw we can define it as $d_o = \sum_{i=-l_d}^{l_d} p_i w_i i^2$. Here $p_{\pm i}$ is the microscopic transition probabilities associated with the forward and reverse movements of 1Drw along with the corresponding microscopic transition rates $w_{\pm i}$ and $i$ is the step length measured in bps. For unbiased sliding movement $p_{\pm i} \sim \frac{1}{2}$ and $w_{\pm i} \sim 10^6$ s-1 so that $d_o \sim 10^6$ bps$^2$ s$^{-1}$. | bps$^2$ s$^{-1}$ |
| $D_o$ | $= d_o/\phi l_d^2$. When the hop size $k > 1$, then $D_o = (k+1)(2k+1)/6$. | dimensionless |
| $u_0$ | Average 1Dd length associated with the dynamics of TFs on DNA. | bps |
| $p_L$ | Probability associated with the entry of TFs from left side of the SBS. | |
| $p_R$ | Probability associated with the entry of TFs from right side of the SBS. | |
| $u$ | Distance between the adjacently located CRMs. | bps |
| $p(u)$ | Probability associated with the TFs which landed at the interconnecting region between two adjacently located CRMs to reach any of them without dissociation. | |
| $q$ | Length of genomic DNA spanned upon binding of TFs of interest. | bps |
| $q_C$ | (Solution of $(N-q)(1/4)^q = 1$ for $q$) critical length of binding stretch of TFs at which only one of the SBS can be found on the genome of size $N$ bps. | bps |
| $p_{LZ}$ | $(=1-X_Z/X_R)$ splitting probability associated with the 1Drw to reach $X_L = 0$ starting from $X_Z$ when $X_A = X_R$. | |
| $p_{Lr}$ | $(=1-X_r/X_R)$ splitting probability associated with the 1Drw to reach $X_L = 0$ starting from $X_r$ when $X_A = X_R$. | |